\documentclass[aps,twocolumn,showpacs,amsmath,amssymb,10pt]{revtex4}  

\begin{document}
\title{On one parametrization of Kobayashi-Maskawa matrix}
\author{Petre Di\c t\u a} \email{dita@zeus.theory.nipne.ro}
\affiliation{Institute of Physics and Nuclear Engineering, P.O. Box MG6, Bucharest, Romania}
\date{December 2, 2009}

\begin{abstract}
An analysis of  Wolfenstein parametrization for the Kobayashi-Maskawa matrix shows that it has a serious flaw: it depends on {\em three} independent parameters instead of {\em four} as it should be.  Because this approximation is currently used in phenomenological analyzes from the quark sector, the reliability  of almost all phenomenological results is called in  question. Such an example is the latest PDG fit from \cite{CA}, p. 150. The parametrization cannot be fixed since even when it is brought  to an exact form it has the same flaw and its use lead to many inconsistencies.
\end{abstract}


\maketitle

 At the beginning of  '70s  Kobayashi and Maskawa (KM), \cite{KM}, have  introduced the six quark model by making   the observation that in such a model the unitary matrix representing charged weak currents has one phase parameter in addition to the real mixing angles. This extra phase introduces  $CP$ violation in a natural way as a result of weak mixing between the quarks, and   experiments at the mid of '70s have shown that in addition to the light quarks that make up ordinary hadrons, there is a charmed heavy quark.
It was easily observed that the anomaly cancellation, true and delicate in the four-quark model, can be restored in the Weinberg-Salam model if there are two more quarks, t and b. In this way the first pillar of the future Standard Model was built. 

A decade latter Wolfenstein,  \cite{W}, used an other form for the KM matrix, \cite{KM}, the so called Murnagham form, \cite {Mur}, and  this one is still  used today by the flavor community, see   \cite{CA}, which is 

\begin{eqnarray}
U=~~~~~~~~~~~~~~~~~~~~~~~~~~~~~~~~~~~~~~~~~~~~~~~~~~~~~~~~~~~~~~~~~\label{ckm}\\
{\small \left[\begin{array}{ccc}
c_{12}c_{13}&c_{13}s_{12}&s_{13} e^{-i \delta}\\
-c_{23}s_{12}-c_{12}s_{23}s_{13}e^{i \delta}&c_{12}c_{23}-s_{12}s_{23}s_{13}e^{i \delta}&s_{23}c_{13}\\
s_{12}s_{23}-c_{12}c_{23}s_{13}e^{i \delta}&-c_{12}s_{23}-s_{12}c_{23}s_{13}e^{i \delta}&c_{23}c_{13}
\end{array}\right]}\nonumber
\end{eqnarray}
Even if Maiani and other theorists shown us that all possible phases from the first row and last column can be eliminated since all the quark fields are not measurable quantities, \cite{M}, the nowadays form  contains a (unobservable) phase for the $b-$quark field, which coincides with the $CP$-violation phase, see Eq.(\ref{ckm}).

For a better understanding of  what Wolfenstein did we start with his footnote no 3 where he says: {\it My notation is more closely related to that of L. Maiani}, \cite{M}. The KM matrix mixing  parameters have been  written by Wolfentsein as 
\begin{eqnarray}s_{\theta}=\lambda,\quad s_{\gamma}=\lambda^2 A,\quad{\it and}\quad s_{\beta}e^{-i \delta}=\lambda^3 A(\rho-i\eta)\label{wolf}\end{eqnarray}
The quoted statement is not true because $ s_{\beta} $ in  Maiani form of KM matrix has no exponential factor, see  \cite{M}.
 The present day notation for mixing angles is
\begin{eqnarray} s_{12}=s_{\theta}, \quad s_{23}= s_{\gamma}\quad {\rm and}\quad  s_{13}=s_{\beta}\label{m1}\end{eqnarray}

Wolfenstein gave an approximate form of KM matrix \cite{KM} by an expansion in parameter $\lambda$ that was considered small enough.

The first remark is  that $\lambda$ enters in the definition of all mixing angles, see (\ref{wolf}), fact that implies that there is a close relationship between them, even if they are considered independent parameters in matrix (\ref{ckm}).

 This ``kinship'' can be obtained from  paper \cite{SS}, whose authors had the  idea to use the exact form of KM matrix which follows by using relations (\ref{wolf}) in the  form (\ref{ckm}). Thus from  
equations (\ref{wolf}) they obtained the following relations
\begin{eqnarray}
\rho=\frac{s_{13}}{s_{12} s_{23}}\cos\delta, \quad \eta=\frac{s_{13}}{s_{12} s_{23}}\sin\delta
\label{ss}\end{eqnarray}
see their formulae (2a)-(2b) in \cite{SS}, which are our starting point. The preceding two equations are equivalent to the following two
\begin{eqnarray}
\rho^2+\eta^2=\left(\frac{s_{13}}{s_{12} s_{23}}\right)^2,\quad\tan\delta =\frac{\eta}{\rho} \label{wr}\end{eqnarray}
Our second remark  is that the first  relation (\ref{wr}) shows that $\rho$ and $\eta$  are not independent parameters because all the mixing angles $s_{ij}$ are independent. That means that the  matrix obtained by the substitution of formulae (\ref{wolf})  into the KM matrix form (\ref{ckm}) leads to a matrix parametrized by {\em three} independent parameters $\lambda,\,A$, and either $\rho$, or $\eta$, instead of {\em four}, as it should be. In other words the relations (\ref{wolf}) do not give a one-to-one transformation between the parameters $s_{ij}$ and $\delta$ entering (\ref{ckm}), and $\lambda,\,A,\,\rho$, and $\eta$ enetring (\ref{wolf}). By  consequence, even if the resulting matrix could be unitary, the above fact  implies that  there is an entire class of   matrices which cannot be recovered from experimental data when using the last group of parameters, and nobody did not estimate the systematic error implied by such a parametrization.

 To see that the above parametrization is  flawed we adhere to Jarlskog's point of view which consists in determination of quark mixing matrix in terms of measurable invariants. Two of them are the KM matrix moduli and the celebrated $J$ invariant, see \cite{J}. 

It is well known that KM moduli enter in leptonic and semileptonic decays
through  products of the form $|U_{pp'}|f_P $ and  $|U_{pp'}f_+(q^2)| $, the first in leptonic, and the second in semileptonic decays. Thus the physics reality suggests the use of KM moduli in any phenomenological analysis from flavor sector, \cite{JS}, since the mixing angles are not measurable quantities. On the other hand by using KM moduli as parameters, instead of mixing angles, we lost the unitarity property of KM matrix (\ref{ckm}). Hence an important problem that has to be solved is the consistency  problem  with unitarity of the KM moduli which amounts to find the necessary and sufficient conditions on the set of numbers $V_{ij}=|U_{ij}|$ for this set to represent the moduli of an exact unitary matrix. This problem was solved in \cite{Di}. The new form of unitarity constraints says that the four independent parameters $s_{ij}$ and $\delta$ should take physical values, i.e. $s_{ij}\in(0,1)$, and $\cos\delta\in(-1,1)$, when they are computed via equations set:

\begin{eqnarray}
V_{ud}^2&=&c^2_{12} c^2_{13},\,\, V_{us}^2=s^2_{12}c^2_{13},\,\,V_{ub}^2=s^2_{13}\nonumber \\
 V_{cb}^2&=&s^2_{23} c^2_{13},\,\,
 V_{tb}^2=c^2_{13} c^2_{23},\nonumber\\
V_{cd}^2&=&s^2_{12} c^2_{23}+s^2_{13} s^2_{23} c^2_{12}+2 s_{12}s_{13}s_{23}c_{12}c_{23}\cos\delta,\nonumber\\
V_{cs}^2&=&c^2_{12} c^2_{23}+s^2_{12} s^2_{13} s^2_{23}-2 s_{12}s_{13}s_{23}c_{12}c_{23}\cos\delta,~~~~~\label{unitary}\\
V_{td}^2&=&s^2_{13}c^2_{12}c^2_{23}+s^2_{12}s^2_{23}-2 s_{12}s_{13}s_{23}c_{12}c_{23}\cos\delta\nonumber,\\
V_{ts}^2&=&s^2_{12} s^2_{13} c^2_{23}+c^2_{12}s^2_{23} +2 s_{12}s_{13}s_{23}c_{12}c_{23}\cos\delta\nonumber
\end{eqnarray}
relations easily obtained from (\ref{ckm}).

 Now we choose as independent parameters the moduli $V_{us}\,V_{ub},\,V_{cb},$ directly related to  $\lambda,\,A,\,\rho,\,\eta$, and $V_{cd}$. 
With the above notations we find from relations  (\ref{unitary}) 
\begin{eqnarray}
s_{12}=\frac{V_{us}}{\sqrt{1-V_{ub}^2}},\,\,
  s_{23}=\frac{V_{cb}}{\sqrt{1-V_{ub}^2}},\,\,
 s_{13}=V_{ub}\label{coef}
\end{eqnarray}
 relations  which are used in (\ref{wr}) to get 
\begin{eqnarray}
\rho^2+\eta^2=\left(\frac{V_{ub}(1-V_{ub}^2)}{V_{us} V_{cb}}\right)^2\label{nu}\end{eqnarray}
relation that shows again that $\rho$ and $\eta$ are not independent parameters.  The above  relations, (\ref{coef}) and (\ref{nu}), show  that all the  parameters
 $\lambda,\,A,\,\rho,\,\eta$ depend on three independent moduli.

To compare the two approaches we consider first and ``academic'' problem, the reconstruction of a unitary matrix when we know exactly all the $U$ matrix moduli, and the simplest case is that of equal moduli $V_{ij}^2=1/3, \,\,i,\,j=1,2,3$, which are the moduli of the  $3\times 3$ unitary Fourier matrix, such that all the computations can be done by hand. In this case the standard unitarity triangle, that is used in almost all phenomenological analyses, is an {\it equilateral} triangle, and as usual its lengths could be taken equal to unity. Then
\begin{eqnarray}
\rho=\frac{1}{2},\quad \eta=\frac{\sqrt{3}}{2} \label{r2} \end{eqnarray}
and from second Eq. (\ref{wr}) we get
\begin{eqnarray}
\tan\delta=\frac{\eta}{\rho}=\sqrt{3}, \quad \delta=\frac{\pi}{3}=60^{\circ} \label{r3}\end{eqnarray}

From relations (\ref{coef}) we find
\begin{eqnarray}
s_{13}=\frac{1}{\sqrt{3}},\,\;\;\; s_{12}=s_{23}=\frac{1}{\sqrt{2}}\label{fou}\end{eqnarray}
Because all the parameters entering (\ref{ckm}) are determined, see relations (\ref{r2})-(\ref{fou}), we can use them for the determination of all KM moduli for this academic case. Using them in Eqs. (\ref{unitary}) we found 
\begin{eqnarray}
V_{ud}^2&=& V_{us}^2=V_{ub}^2=
 V_{cb}^2=
 V_{tb}^2=\frac{1}{3}\nonumber\\
V_{cd}^2&=&V_{ts}^2=\frac{1}{3}+\frac{\sqrt{3}}{12}\\
V_{cs}^2&=&V_{td}^2=\frac{1}{3}-\frac{\sqrt{3}}{12}\nonumber
\end{eqnarray}
From the above numerical results on can see that  by using  Wolfenstein parametrization one cannot recover from ``data'' the simplest unitary matrix even when the parametrization is an {\it exact} one, and  the above numbers represent the simplest test proving its flaw.

In our approach the mixing angles take the same values, (\ref{fou}), and from the sixth relation (\ref{unitary}) we find 
\begin{eqnarray}
 \cos\delta=0,\;\;\;\delta=\pm \frac{\pi}{2}
 \label{r1}\end{eqnarray}
The above relation shows that we have an ambiguity in choosing $\delta$, and to resolve it  our choice is Im$U_{21} >0$, that implies Im$U_{22} <0$, Im$U_{31} <0$, and  Im$U_{22} >0$. However  we must take into account that $U_{13}$ is complex. By consequence we have to multiply (\ref{ckm}) at right by the diagonal matrix
$d_1=(e^{i \delta},1,1)$, followed by a second diagonal matrix $d_2=(1,e^{-i \delta}e^{-i \delta},)$ to bring it at its rephaised form, such that, for example,  $U_{21}$ has the form
\begin{eqnarray}
U_{21}=-c_{23}s_{12}e^{-i \delta}-c_{12}s_{23}s_{13}
\end{eqnarray}
In our case  Im$U_{21}$ is positive  when $\delta=\frac{\pi}{2}$, and we get
\begin{eqnarray}
U_{21}&=&-\frac{1}{2\sqrt{3}}+\frac{i}{2}= e^{2\pi i/3},\\\nonumber
U_{22}&=&-\frac{1}{2\sqrt{3}}-\frac{i}{2}= e^{4\pi i/3},\;{\rm etc}
\end{eqnarray}
 recovering in this way the known form, up to equivalence, of the 3-dimensional Fourier matrix. 

The big difference between $\delta=\frac{\pi}{3}$, obtained from the second Eq. (\ref{wr}), and the true value $\delta=\frac{\pi}{2}$, shows that Wolfenstein parametrization is senseless, and the flavor community has to give up it.

Up to now we discussed an exact academic example. In the following we show that the central values of KM moduli matrix given in \cite{CA}, p. 150, are not compatible with unitarity constraints. For proving that we need to put into game another independent modulus. If this is $V_{cd}$ from the sixth equation (\ref{unitary}) one gets
\begin{widetext}
\begin{eqnarray}
\cos\delta = 
\frac{(1-b^2)(V_{cd}^2(1-b^2)-a^2)+c^2(a^2+b^2(a^2+b^2-1))}{2 a b c \sqrt{1-a^2-b^2}\sqrt{1-b^2-c^2}\label{col}}
\end{eqnarray}
\end{widetext}
and three similar formulae from the last three equations. The above relation shows that $\cos\delta$ is an other invariant in the Jarlskog sense that depends on four independent moduli, and $CP$-violation phase can be measured via relations such as (\ref{col}). If one makes use of the last four relations (\ref{unitary}) we get only one solution for the mixing angles and $\cos\delta$. Because there are 58 such groups of four independent moduli one get 165 different formulae for
 $\cos\delta$. They take the same numerical value if and only  if all the six relations similar to 
\begin{eqnarray}
V_{ud}^2+V_{us}^2+V_{ub}^2=1\label{un1}
\end{eqnarray}
are exactly satisfied. If the moduli matrix generated by four independent moduli is compatible with unitarity then $\cos\delta\in (-1,1)$, and outside this interval when the corresponding matrix is not compatible. In the following we work with data from matrix (11.27) from \cite{CA}, p. 150, which has the form (\ref{cls}). In the following we will use all the central values as rational numbers, i.e. $V_{ud}=9741/10^5$,  $V_{us}=2257/10^4$, etc, because the central values from 
(\ref{cls}) does not come from an exact matrix, which means that the six relations similar to (\ref{un1}) take different values, as the following  numerical computations show
\begin{eqnarray}\begin{array}{ccc}
V_{ud}^2+V_{us}^2+V_{ub}^2-1&=&-4.658\times 10^{-7}\\*[2mm]
V_{cd}^2+V_{cs}^2+V_{cb}^2-1&=&~ 8.366\times 10^{-6}\\*[2mm]
V_{td}^2+V_{ts}^2+V_{tb}^2-1&=& -3.707\times 10^{-7}\\*[2mm]
V_{ud}^2+V_{cd}^2+V_{td}^2-1&=&1.79\times 10^{-5}\\*[2mm]
V_{us}^2+V_{cs}^2+V_{ts}^2-1&=&-1.226\times 10^{-5}\\*[2mm]
V_{ub}^2+V_{cb}^2+V_{td}^2-1&=&1.89\times 10^{-6}\end{array}\label{ds}
\end{eqnarray}
results that from a phenomenological point of view could be acceptable.
\begin{widetext}
\begin{eqnarray}
|U|=\left[\begin{array}{ccc}
0.97419\pm0.00022&0.2257\pm0.0010&0.00359\pm0.00016\\*[2mm]
0.2256\pm0.0010&0.97334\pm0.00023&0.0415^{+0.0010}_{-0.0011}\\*[2mm]
0.00874^{+0.00016}_{-0.00037}&0.0407\pm0.0010&0.999133^{+0.000044}_{-0.000043}\label{cls}\end{array}\right]
\end{eqnarray}\end{widetext}
 If we use all the 165 different formulae for $\cos\delta$ one get  from (\ref{cls})
\begin{eqnarray}
\langle\cos\delta\rangle=0.467-0.005\, i\nonumber\\
\sigma_{\langle\cos\delta\rangle}=0.245+0.0096\, i\label{ds1}
\end{eqnarray}
where $i=\sqrt{-1}$ is the imaginary unit. The above result shows that the central values of matrix (\ref{cls}) are not compatible with unitarity.
 The complex values in (\ref{ds1}) come from the matrix determined by the following four independent central values:
\begin{eqnarray}V_{ud}=\frac{97419}{10^5},\;V_{cd}=\frac{2256}{10^4},\;V_{cb}=\frac{415}{10^4},\; V_{ts}=\frac{407}{10^4} \label{ds2}\end{eqnarray} 
If we use  relations similar to (\ref{un1}) to find all the entries of the corresponding KM moduli matrix, we find, e.g., $V_{ub}^2=1-V_{ud}^2-V_{cd}^2-V_{cb}^2+V_{ts}^2$, etc. If we compute the previous expression with the numerical values (\ref{ds2}) we get $V_{ub}^2= -72761/10^{10}$, i.e. $V_{ub}=0.0027\,i$. If we compare with the corresponding values from (\ref{cls}) we see that both of them are of the same order of magnitude, but the second one is an imaginary quantity. Thus the $V_{ub}$ error from (\ref{cls}) is meaningless, and the  central values matrix (\ref{cls}) do not satisfy all  unitarity constraints. As a matter of fact this is a general phenomenon. For example if we add to $V_{us}$ modulus the small value $3
\times 10^{-4}$, that is  smaller than the error $10^{-3}$, see (\ref{cls}), and we use the independent moduli $V_{us},\;V_{ub},\;V_{cd},\;V_{cb} $ we find
\begin{eqnarray}V_{td}=\sqrt{V_{us}^2+V_{ub}^2-V_{cd}^2}= \frac{i\sqrt{503717119}}{10^5}\approx 0.224\,i\end{eqnarray}
 Now we compute $\cos\delta$  from all formulae such as (\ref{col}) with the numerical values from (\ref{ds2}),  and  we find the same value because the corresponding matrix is the square root of an exact double stochastic matrix, see \cite{Di}, and the numerical result is the following
\begin{widetext}\begin{eqnarray}
\cos\delta=-\frac{820129143378686555297\sqrt{13}\,i}{44664292\sqrt{21160837372396659}\times 10^6}\approx-0.455\,i\label{ms}\end{eqnarray} 
If we make use of the central values $V_{us}=2257/10^4,\;V_{ub}=359/10^5,\,V_{cd}=2256/10^4,\; {\rm and} \; V_{cb}=415/10^4$ we obtain
\begin{eqnarray}
\cos\delta=\frac{11331307828401955126433}{1817617\sqrt{94739989494766501561}\times 10^6}\approx 0.640 \label{ds3}\end{eqnarray}
\end{widetext}
Thus the central values matrix (\ref{cls}) is not unitary because not all $\cos\delta$  take physical values, i.e.  $\cos\delta\in (-1,1)$, and they do not satisfy the physical conditions  $\cos\delta^{(i)}\approx \cos\delta^{(j)},\;\;i\ne j$. As we have shown in paper \cite{Di} the $\chi^2$-function must have two contributions. The first one
\begin{eqnarray}\chi^2_{1}&=&
\sum_{j=u,c,t}\left(
\sum_{i=d,s,b}V_{ji}^2-1\right)^2
+\sum_{j=d,s,b}\left(
\sum_{i=u,c,t}V_{ij}^2-1\right)^2\nonumber\\
&&+ \sum_{i < j}(\cos\delta^{(i)} -\cos\delta^{(j)})^2,\,\,\,\,-1\le\cos\delta^{(i)}\le 1 \label{chi1}
\end{eqnarray}
which enforces all the unitarity constraints, and a second one that take into account the experimental data under the form 
\begin{eqnarray}
\chi^2_2=\sum_{i}\left(\frac{d_{i}-\widetilde{d}_{i}}{\sigma_{i}}\right)^2\label{chi2} 
\end{eqnarray}
where  $d_{i}$ are the theoretical functions one wants to be found from fit, $\widetilde{d}_{i}$ are  the numerical values that describe the corresponding experimental data, while  $\sigma_i$ is the  uncertainty  associated to $\widetilde{d}_{i}$.

In conclusion the above numerical computations obtained by using  the central values from the last PDG fit in paper \cite{CA}, p.150, show that the Wolfenstein parametrization is wrong even when it is made exact, and must to be abandoned if we want to obtain physical reliable numbers for flavor physics parameters.
\vskip3mm
{\bf Acknowledgements.}  We acknowledge a partial support from  ANCS contract no 15EU/0604.2009.

\end{document}